# The Fractured Boer-Mulders Effect in the Production of Polarized Baryons

Dennis Sivers
Portland Physics Institute
4730 SW Macadam Ave. #101
Portland, OR 97239

Spin Physics Center
University of Michigan
Ann Arbor, MI 48109-1040

**Abstract**

The fractured Boer-Mulders functions, $\Delta^N M^q_{B\uparrow/\{q,q\}\uparrow:p}(x, p_{TN}; z, \vec{p}_T \cdot \vec{k}_T; Q^2)$, describe an intriguing class of polarization effects for the production of baryons in the target fragmentation region of deep-inelastic processes. These functions characterize transverse momentum asymmetries related to the spin orientation for different flavors of axial-vector diquarks, $\{q_i, q_j\}\uparrow$, in an unpolarized ensemble of protons just as the familiar Boer-Mulders functions characterize transverse momentum asymmetries connected to the spin orientation of quarks in unpolarized targets. The asymmetries in $p_{TN}$ of the fractured Boer-Mulders effect originating in the proton distribution function can be separated kinematically, both in SIDIS and in the Drell-Yan process, from the asymmetries in $k_{TN}$ of the polarizing fracture functions, $\Delta^N M^q_{B\uparrow/(q,q):p}(x, p_T^2; z, k_{TN}; Q^2)$, generated during the soft color rearrangement of the fragmentation process. The experimental requirements for this separation are presented in this article and it is shown that the fractured Boer-Mulders effect should change sign between Drell-Yan and SIDIS while the polarizing fracture functions remain the same. Simple isospin arguments indicate the two polarization mechanisms should give significantly different results for the production of polarized $\Lambda$'s and $\Sigma$'s.

## I. Introduction

The experimental results on the production of transversely polarized hyperons in hadron-hadron collisions [1,2] have played an extremely important role in motivating the early development of a systematic theoretical approach to parity-conserving single-spin observables in high-energy processes. Based on the null result for single-spin asymmetries in perturbative QCD found by Kane, Pumplin and Repko [3], it can be explicitly demonstrated [4] that all such polarization observables must be associated with coherent spin-orbit dynamics in the nonperturbative sector of the theory. While the overall implications of this important theoretical result for baryon polarization asymmetries have long been apparent, the phenomenological details specific to this particular application have not been directly pursued. Some of the reasons for this type of neglect can be attributed to the manner in which the relevant theoretical concepts have been framed. The coherent mechanisms leading to single-spin observables can be parameterized as twist-3 operators [5,6] in a collinear factorization of the hard-scattering model of QCD. Equivalently, they can be described in terms of particular $k_T$-dependent parton distribution functions or parton fragmentation functions [7,8]. Like the careful gestures of a stage magician, both of these formalisms direct primary attention to asymmetries found in the "current fragmentation region" of deep-inelastic processes. Spin asymmetries found in baryon production, however, require that we look more carefully at dynamical mechanisms found in the target fragmentation region of such processes.

The change in kinematic venues requires some additional tools and it is helpful to review the existing conventions in order to formulate a congruent approach. Mulders and Tangerman [9] have identified four classes of "leading-twist" quantum structures resulting in single spin observables that occur in the formulation of a $k_T$- dependent extension of the operator product expansion. This quartet of functions is presented in Table I. Parity conserving single-spin observables are odd under a particular symmetry, here designated $A_\tau$, constructed with the aid of the Hodge dual operator of differential geometry. The Mulders-Tangerman classification identifies two sets of $A_\tau$-odd fragmentation functions, the Collins functions [10] and the polarizing fragmentation functions [11]. The $A_\tau$-odd parton distribution functions classified here are the Boer-Mulders functions [12] and the orbital distribution functions [4]. Table I gives the relationship between the expressions for partonic number densities used by the author and the expressions used for the related partonic correlators that have the dimension of an inverse momentum. The conventions for the signs and the factors of 2 found in this table are explained in the article by A. Bacceta, U. D'Alesio, M. Diehl and C.A. Miller [13] that defines the "Trento Conventions" for this field. The scalar quantity

$$k_{TN} = \vec{k}_T \cdot (\hat{\sigma} \times \hat{p}) \qquad (1.1)$$

found in this table defines an, $A_\tau$-odd and P-even, spin-directed momentum which is characteristic of every transverse single-spin observable. The quantitative value of the concept of spin-directed momentum is discussed more extensively in [14,15]. The identification of the mechanism or mechanisms leading to a spin-directed momentum responsible for a single-spin observable plays an important role in the analysis presented here. As indicated above, a crucial component in this identification takes advantage of KPR factorization. The concept of KPR factorization applies the result of ref. [3] which demonstrates that single-spin asymmetries involving light quarks in perturbative QCD are tiny to incorporate the coherent dynamics leading to a spin-directed momentum either into an effective "distribution function" for an initial-state particle or into the coherent color rearrangement leading to the "fragmentation" of a color constituent producing a final-state hadron.

The concept of KPR factorization embodied in the operators of Table I has proved to be extremely useful for phenomenology in the current fragmentation region. Extensive study of experimental results for single-spin asymmetries of $\pi$'s and $K$'s produced in semi-inclusive deep inelastic scattering (SIDIS) [16,17,18] and in pp collisions [19, 20] has led to significant advances in the understanding of proton spin structure and nonperturbative processes in QCD. Notable successes include the separation of asymmetries associated with the Collins-Heppelmann [21] mechanism from those resulting from orbital distributions. Combining the results on the Collins-Heppelmann effect with the extraction of Collins functions from analysis of $e^+e^-$ data. [22] then leads to a consistent specification of the quark transversity distributions. [23]

A convenient method for extending the phenomenological conventions to allow the inclusion of polarized hyperon production into this very successful framework for single-spin asymmetries can be found in the formalism for <u>fracture functions</u>, $M^b_{h/p}(x, z_h; Q^2)$, introduced by Veneziano and Trentadue [24]. The appellation "fracture" was chosen by those authors to suggest a hybrid form combining fragmentation functions and structure functions. These fracture functions characterize the conjoint probability for finding both the parton b (quark or gluon) with Bjorken $x = Q^2 / 2 p.q$, and the hadron h (meson or baryon) with Feynman $z_h = p_h \cdot p / q \cdot p$ in a semi-inclusive deep inelastic process involving a proton target. Conjoint probability distributions are powerful tools in the study of quantum mechanical systems and including both the spin designation and the specification of transverse momentum asymmetries into the fracture function formalism then provides a flexible approach to characterizing the dynamics involved in single-spin observables. The fracture function $M^u_{\pi/p\uparrow}(x,\vec{p}_T;z,\vec{k}_T;Q^2)$, for example, describes the final state in lepton scattering from a transversely polarized proton target in which the jet of the u quark has transverse momentum $\vec{p}_T$ and the detected $\pi$ has transverse momentum $\vec{k}_T$ in a reference frame defined by $\vec{q}=Q\hat{e}_z$. [25] The classifications of fracture functions describing the $k_{TN}$-odd asymmetries in SIDIS are shown in Table II. Final-state hadrons in the fracture function formalism can occur either in the current fragmentation region, the target fragmentation region or in the central "plateau" region that connects the other two. For hadrons produced in the current fragmentation region, the functions in Table II can thus replicate the existing phenomenology on transverse spin asymmetries. For example the Collins-Heppelmann fracture functions, $\Delta^N M^{q\uparrow}_{h/p\uparrow}(x,p_T^2;z_h,k_{TN};Q^2)$, encode the description of a chiral-odd quark transversity distribution convoluted with a chiral-odd and $A_\tau$-odd Collins function as found in the familiar Collins-Heppelmann mechanism without necessarily requiring hard-scattering factorization. [21]

Extending these functions to consider the production of baryons in the target fragmentation region, [26] is then a natural augmentation of the fracture-function formalism. This generalization does require, however, a nontrivial modification involving the inference of quantum structures describing diquarks in scattering processes. This issue has significant impact on understanding how to extend three of the original dynamical mechanisms in the Mulders-Tangerman classification to describe spin asymmetries in baryon production. Diquarks are not fundamental constituents but are themselves composite systems representing correlations generated in the nonperturbative regime of QCD by the confinement of color in baryons. The distinction between pointlike constituents and localized correlations can be partially evaded in the treatment of polarizing fracture functions found in ref. [26]. However, extending the Collins-Heppelmann mechanism, the Boer-Mulders effect and the orbital effect to the production of baryons in the target fragmentation region directly confronts the fact that the resulting functions do not parameterize asymmetries in the spin-directed momenta of the quarks undergoing the hard scattering, but instead, must describe $A_\tau$-odd asymmetries involving the composite diquark correlations in the remnant systems left behind in the target fragmentation region. To emphasize this shift in dynamical focus involving the inclusion of composite constituent systems, the author prefers a corresponding shift in nomenclature to describe spin-dependent asymmetries for baryon production in the target fragmentation region. In this work, the term "fractured" is used to designate mechanisms where probability distributions involving diquarks are inferred from the conjoint probabilities of the fracture function formalism for the hard-scattering process. The language suggested in Table III pays full homage to the underlying versatility of the fracture function formalism but respects the clarity implied by the mechanisms in Table I. Thus, the Collins-Heppelmann fracture functions of Table II, $\Delta^N M^{q\uparrow}_{h/p\uparrow}(x,p_T^2;z,k_{TN};Q^2)$, can describe the original Collins-Heppelmann processes in the current fragmentation region of SIDIS but when we consider the "fractured Collins-Heppelmann mechanism", described by the functions $\Delta^N M^q_{B/\{q,q\}\uparrow:p\uparrow}(x, p_T^2; z_B, k_{TN}; Q^2)$, we employ a notation that characterizes an effective "transversity-type distribution" for $J^P = 1^+$ diquarks in a polarized proton that is convoluted with a Collins-type, $A_\tau$-odd, function describing an asymmetry in the fragmentation process for such systems of polarized diquarks. Note that the notation in Table III represents a slight variation of that used in ref. [26] for the same structures. Details outlining the content of this notation will be discussed later in this article.

While it is important to be aware of the larger framework that allows a complete classification for all single-spin asymmetries for baryon production in the target fragmentation region, the focus of this paper is specifically directed at the fractured Boer-Mulders mechanism. The fractured Boer-Mulders functions, written here as $\Delta^N M^q_{B\uparrow/\{q,q\}\uparrow:p}(x, p_{TN}; z_B, \vec{p}_T \cdot \vec{k}_T; Q^2)$, describe an asymmetry in the production of polarized baryons in the target fragmentation region of SIDIS resulting from an $A_\tau$-odd correlation between the spin orientation of $J^P = 1^+$ diquarks and their transverse momentum in an unpolarized ensemble of target protons. The production of polarized baryons from this mechanism then requires an $A_\tau$-even fragmentation process involving these polarized diquark systems. The fractured Boer-Mulders mechanism can thus be combined with the polarizing fracture functions [26] of Table III to account for all the possible leading-twist processes in QCD that can lead to polarization asymmetries in baryon production in the target fragmentation region of hard scattering. There already exists indirect evidence for the fractured Boer-Mulders mechanism from data on polarized $\Lambda\uparrow$ and $\Sigma\uparrow$ production in pp and pA collisions. In single-particle inclusive processes in hadron-hadron collisions it is not possible to distinguish kinematically the fractured Boer-Mulders mechanism from the polarizing fracture functions, $\Delta^N M^q_{B\uparrow/(q,q):p}(x, p_T^2; z_B, k_{TN}; Q^2)$, that are described in Ref. [26]. However, it can easily be demonstrated that the polarizing fracture functions lead to

$$P(pp \to \Sigma_o \uparrow X) \cong -\frac{1}{3} P(pp \to \Lambda_o \uparrow X) \tag{1.2}$$

This result has also long been known to be a consequence of Lund-model [27] phenomenology. Existing experimental results [28], however, find the result

$$P(pp \to \Lambda_o \uparrow X) \cong -P(pp \to \Sigma_o \uparrow X) \tag{1.3}$$

Allowing for the impact of small effects associated with the decay of higher-mass resonances, this large discrepancy between experiment and the Lund model strongly indicates the need for a polarization mechanism for these two hyperons that is <u>not</u> part of the fragmentation process. Since the mechanism must therefore involve an effective diquark distribution, it is likely that the fractured Boer-Mulders mechanism has already been observed! Unlike the polarizing fracture functions, full calculations for the fractured Boer-Mulders functions necessarily involve intricacies associated with initial-state or final-state interactions that preclude unambiguous predictions. However, the sign of the polarization in the fractured Boer-Mulders effect can be inferred from simple kinematic arguments and much can be learned from isospin constraints. It turns out that the comparison of the fractured Boer-Mulders functions for the two hyperon polarization processes $pp \to \Lambda_o \uparrow X$ and $pp \to \Sigma_o \uparrow X$ can provide a good starting point for the discussion of diquark structure in the proton. A nontrivial experimental test of the underlying ideas is that our approach predicts that the fractured Boer-Mulders mechanism should change sign when comparing polarization asymmetries of baryons produced in SIDIS with those found in associated production of the Drell-Yan process. This important new result follows from the same Collins conjugation arguments [29] leading to predictions for the change in sign of orbital distributions in SIDIS and Drell Yan that have already received considerable attention. The new predictions for baryon polarization asymmetries can be tested without requiring polarized beams or targets.

The remainder of this paper is organized as follows. Section II discusses the origin of the fractured Boer-Mulders effect in terms of a spin-directed momentum transfer associated with virtual fluctuations involving axial vector diquarks. Section III describes the parallelism conjecture relating the initial-state or final-state interactions found in Boer-Mulders functions with those found in orbital distributions. This connection naturally extends to the fractured version of these two dynamical structures. Section IV contains the main result of the paper with specific predictions for the polarization of $\Lambda$'s and $\Sigma$'s produced in the target fragmentation region of SIDIS or of the Drell-Yan process. Section V then presents some preliminary ideas concerning target spin asymmetries in baryon productions processes.

## II. Fractured Boer-Mulders and Diquarks

As specified in the introduction, the fractured Boer-Mulders mechanism involves an $A_\tau$-odd correlation between the transverse momentum distribution of $J^P = 1^+$ diquarks in the target fragmentation region of a hard-scattering event involving an unpolarized ensemble of target nucleons and the spin orientation of these axial vector diquarks. Like other dynamical mechanisms leading to transverse single-spin asymmetries, the genesis of the fractured Boer-Mulders mechanism involves spin-orbit correlations. To illustrate how these correlations can lead to asymmetries in the production of polarized baryons, it is helpful to first specify what we mean by the transverse momentum distribution of diquarks.

For this paper, we will describe diquarks within the framework of a broken $SU(3)_f$ of flavor and the confining $SU(3)_c$ of QCD. The minimum energy configuration for localized two-quark system is a color antitriplet and this color configuration in a color-singlet baryon leads to the flavor antitriplet of $J^P = 0^+$ diquarks that we will designate

$$[q,q] = [u,d],[d,s],[s,u] \qquad (2.1)$$

and to the flavor sextet of $J^P = 1^+$ diquarks whose flavor content is given by

$$\{q,q\} = \{d,d\},\{u,d\},\{u,u\},\{d,s\},\{s,u\},\{s,s\}. \qquad (2.2)$$

As suggested by the nomenclature, the $[q,q]$ diquark configurations are antisymmetric under the interchange of quarks while the $\{q,q\}$ configurations are symmetric. Further detailed information concerning such diquark composite systems can be found in the comprehensive articles by Jaffe [30]. These articles include an operator description of such localized correlations. For convenience, we will frequently refer to the angular momentum content of a diquark system as "spin" suggesting an underlying pointlike interpretation. When the angular-momentum orientation of the $J^P = 1^+$ configuration is involved we will use the designation $\{q,q\}\uparrow$. To simplify expressions when writing equations that apply for both scalar and axial-vector diquarks, we will use the notation $(q,q)$ to describe any color antitriplet diquark. Feynman diagram calculations involving simple quark-diquark models for the proton [31] have played an important role in the construction of phenomenological expressions for orbital distributions and other $A_\tau$-odd quantum structures. However, this study is not specifically considering only such diquark models. The basic assumption underlying the treatment of diquarks in this article is that they represent "localized' rather than "local" structures and that are dynamically important in baryon production processes because diquarks retain their identity in soft, nonperturbative interactions involving other color-charged gluonic systems in QCD. The retention of isospin and spin quantum numbers occurs because interactions involving a flavor interchange with another quark in the system are required to alter the flavor symmetry properties of diquarks. The localized nature of diquarks implies that we can use the fracture function approach to infer an effective distribution of momentum for them in a hard-scattering event. After the hard scattering of a quark with initial longitudinal momentum determined by $x = Q^2 / 2p \cdot q$ and transverse momentum $\vec{p}_T$, the fracture function $M_{h/p}^q(x, \vec{p}_T; z, \vec{k}; Q^2)$ can be used to infer an effective distribution both for the scattered quark

$$f_{eff}^q(x, \vec{p}_T) \qquad (2.3)$$

and for the target remnant system

$$f_{eff}^{rem}(x^{rem}, p_T^{rem}), \qquad (2.4)$$

where, at this stage, we can assume a "quasi-elastic" relationship between the kinematics of the target remnant system and the scattered quark,

$$\begin{aligned} x^{rem} &\cong 1 - x \\ \vec{p}_T^{\,rem} &\cong -\vec{p}_T. \end{aligned} \qquad (2.5)$$

We then hypothesize that this target remnant system contains a localized diquark that can subsequently "fragment" by picking up another quark to form a baryon in the final state.

The conditional probability of finding this diquark in the target remnant system thus defines an effective diquark distribution

$$f_{eff}^{(q,q)}(x', \vec{p}_T')  \tag{2.6}$$

that can then be used to characterize initial aspects of the fragmentation process that leads to a baryon in the final state.

To simplify the discussion we will ignore correlations with the longitudinal momentum variables and concentrate specifically on the processes that can generate a spin-directed transverse momentum. Consider, therefore, the transverse momentum of this baryon produced in the target fragmentation region of a semi-inclusive deep-inelastic lepton scattering process as indicated in Fig. 1. Contributions to the baryon's transverse momentum can be separated into three categories:

1. Initial transverse momentum, $\vec{p}_T'$, of the diquark associated with its confinement within the target proton.
2. Transverse momentum, $\vec{Q}_T'$, transferred to the diquark system during the hard-scattering process.
3. Transverse momentum, $\vec{k}_T'$, generated by the soft processes of color rearrangement occurring as the diquark system captures a quark in the fragmentation process to form the polarized baryon.

For the spin-averaged production cross section, the convolution of contributions from these three categories to the transverse momentum of the detected baryon leading to $\vec{k}_T = \vec{p}_T' + \vec{Q}_T' + \vec{k}_T'$ can then be indicated in the form of a diagram such as that shown in Fig. 2. It is understood that the boundaries between these categories are necessarily flexible. The ambiguities arise because the soft or collinear components of the hard-scattering process cannot be assigned uniquely to one category or another. Such an assignment is a matter involving a factorization prescription. [32] One specific feature that can be extracted from Fig. 2 is the observation that the role of diquark transverse momentum in the production of a baryon in the target fragmentation region is quite similar to the role of quark transverse momentum in the production of a meson in the current fragmentation region. For our discussion of baryon polarization asymmetries, we will not be required to explore in detail factorization prescriptions for all such $k_T$-dependent observables [33] because the issue is not the precise definition of the boundaries in Figs. 1 and 2 but the identification of the nonperturbative dynamics of the mechanisms that produce such spin asymmetries. Since the perturbatively-calculable hard-scattering process cannot produce $A_\tau$-odd asymmetries, the mechanisms can therefore be classified according to whether the nonperturbative $A_\tau$-odd dynamics appear in the distribution function of the target or in the fragmentation process..

To discuss these coherent, nonperturbative mechanisms in the production of polarized baryons, it is convenient to explore here in more detail the concept of spin-directed momentum as defined in the introduction. As indicated in the caption of Fig. 1, we specify that the momentum of the detected baryon lies in the x-z plane and that the spacelike momentum transfer, $\vec{q}$, of the lepton scattering process defines the z axis and that the only detected spin observable is given by the polarization vector, $\vec{P} = P\hat{e}_y$, that must be normal to the production plane. The $A_\tau$-odd scalar, $k_{TN}$, defined in Eq. (1.1) can then be written

$$k_{TN} = \vec{k}_T \cdot (\hat{\sigma} \times \hat{q}) = \vec{k}_T \cdot (\hat{e}_y \times \hat{e}_z) = k_x \qquad (2.7)$$

With these directions specified it follows that the $A_\tau$ symmetry projections guarantee that spin density matrices are diagonal in the $\hat{y}$-basis as explained more thoroughly in Ref. [26]. However the factorization prescription is chosen, the spin-directed momentum leading to a polarization asymmetry for the baryon cannot originate in the perturbatively-calculable hard-scattering process. As described in the introduction, this follows from the result of Kane, Pumplin and Repko (KPR) and is designated KPR factorization in Ref. [26]. In this work we are going to extend, based on the explicit calculations of Dharmaratna and Goldstein [34], the concept of KPR factorization to processes involving the production of strange quarks. For fragmentation processes, either the fragmentation of quarks [9,10] or diquarks [26], KPR factorization is consistent with hard-scattering factorization and the $A_\tau$-odd fragmentation functions are process independent. [35] However, for distribution functions such as the Boer-Mulders function considered here, the virtual spin-orbit effects in a stable proton posses a rotation U(1) invariance until a hard scattering occurs. This rotational symmetry leads to the requirement that initial-state and/or final-state interactions involving the hard-scattering process play an essential role in generating the necessary spin-directed momentum. The initial-state and final-state interactions necessarily result in process dependence for an $A_\tau$-odd distribution function. The required process dependence for the spin-directed momentum is easily found and conveniently described in the gauge-link formalism [36]. The resolution of this process dependence with the intrinsic spin-orbit dynamics is described in Refs. [11,14,15]. The result is that hard-scattering kinematics can be used with the "effective" $A_\tau$-odd distribution functions but the functions themselves involve unavoidable process dependence. We will show that this feature is also found in the diquark distribution inferred in the fractured Boer-Mulders effect. We will return to the issue of whether the orbital distributions and Boer-Mulders functions are related to an "intrinsic" property of the target nucleon in Sec. III. The fractured Boer-Mulders effect demonstrates how the coherent spin-orbit effects and initial-state or final-state interactions combine to produce an asymmetry in $p_{TN}$ for an effective distribution of $\{q,q\}\uparrow$ diquarks. Once such an asymmetry exists, it can be convoluted which with an $A_\tau$ even fragmentation function for these diquarks to generate a polarized baryon asymmetry.

With the kinematics specified in Fig. 1, we can then define the fracture function $M^q_{B\uparrow/p}(x, \vec{p}_T; z, k_{TN}\hat{e}_x; Q^2)$ for the production of a polarized baryon and use the fact that the relevant spin density matrices can be diagonalized in the $\hat{y}$ basis to classify the possible sources of a spin-directed momentum in the production process. We define the polarization asymmetry

$$\Delta^N M^q_{B\uparrow/p} = M^q_{B\uparrow/p} - M^q_{B\downarrow/p} \qquad (2.8)$$

and invoke the separation implied by KPR factorization to write

$$\Delta^N M^q_{B\uparrow/p}(x, \vec{p}_T; z, k_{TN}\hat{e}_x; Q^2) = \sum_{(q,q)} \Delta^N M^q_{B\uparrow/(q,q):p}(I) + \sum_{\{q,q\}} \Delta^N M_{B\uparrow/\{q,q\}\uparrow:p}(II) \qquad (2.9)$$

The form of the subscripts in (2.8) can be understood by specifying that the expression $B\uparrow/(q,q):p$ represents the production of a polarized baryon by an $A_\tau$-odd fragmentation process of an unpolarized ensemble of diquarks in the target fragmentation region while $B\uparrow/\{q,q\}\uparrow:p$ represents an $A_\tau$-odd effective distribution of $\{q,q\}\uparrow$ diquarks in the target remnant system followed by an $A_\tau$-even fragmentation of these diquarks into the observed baryons. The separation given by (2.8) can be clarified by the observation that the polarization asymmetry is necessarily an odd function of $k_{TN}$, where

$$\vec{k}_T = k_{TN}\hat{e}_x = (\vec{p}'_T + \vec{Q}'_T + \vec{k}'_T) \qquad (2.10)$$

and the decomposition of transverse momenta is associated with the categories 1.-3. above. The concept of KPR factorization implies that, in any factorization prescription, the $\vec{Q}'_T$ appearing in (2.9) is an even function of $Q_{TN}$. The two remaining possibilities for a spin-directed momentum involve convolutions over the "internal" transverse variables, $\vec{p}'_T, \vec{k}'_T$, that are either:

(I) even in $p'_T \cdot \hat{e}_x = p'_{TN}$ and odd in $\vec{k}'_T \cdot \hat{e}_x = k'_{TN}$

(II) odd in $\vec{p}'_T \cdot \hat{e}_x = p'_{TN}$ and even in $\vec{k}'_T \cdot \hat{e}_x = k'_{TN}$ .

The two allowed possibilities are indicated schematically in Fig. 3. The first possibility is associated with the polarizing fracture(d) functions discussed in Ref. [26] The second distinguishes the fractured Boer-Mulders functions considered here. Because the fractured Boer-Mulders effect involves a soft, coherent spin-directed momentum in an effective distribution of an unpolarized ensemble of protons we can associate the spin orientation it describes with a density matrix for the ensemble of $\{q,q\}\uparrow$, $J^P = 1^+$ diquarks in the target remnant system.

It is important to keep in mind that the specification of an $A_\tau$-odd effective distribution necessarily depends on the existence of initial-state and/or final-state interactions involving the scattering process to break the U(1) symmetry of the spin-directed momentum in a spin-orbit state of a stable composite system. For the fractured Boer-Mulders function, we will see that this breaking of the rotational symmetry is necessarily determined by the geometry of the hard scattering of a quark bound to the diquark in a system where both are orbiting. The interactions resulting from the oriented force between the scattered quark and the $\{q,q\}\uparrow$ system lead to a correlation between the transverse momentum asymmetry of the scattered quark and that of the remnant diquark in a given event of the form

$$\delta p^{,i}_{TN}(diquark) = -(\cos\Theta_i(x_i,...))\delta p^i_{TN}(quark)(1+O(\frac{1}{Q^2})) \qquad (2.11)$$

with $\Theta_i$ a small angle determined by the soft dynamics. Based on (2.11) we can see that an $A_\tau$-odd dynamic effect involving the axial-vector diquark also generates an asymmetry for the scattered quark. This kinematic correlation has the important consequence that it allows the detection of an $A_\tau$-odd quark momentum distribution in the current fragmentation region to select for an $A_\tau$-odd diquark momentum distribution in the target fragmentation region. The connection specified in (2.11) allows for the kinematic distinction between the polarizing fracture(d) functions, $\Delta^N M^q_{B\uparrow/(q,q):p}(x,p_T^2;z,k_{TN};Q^2)$, where the $A_\tau$-odd dynamics occur in the fragmentation process, from the fractured Boer-Mulders functions $\Delta^N M^q_{B\uparrow/\{q,q\}\uparrow:p}(x,p_{TN};z,\vec{p}_T\cdot\vec{k}_T;Q^2)$, where the $A_\tau$-odd dynamics can be absorbed into an effective distribution. The selection is illustrated for the case of semi-inclusive deep inelastic lepton scattering in Fig. 4. The selection is possible because weighting events according to the orientation of the 3 different planes shown in this figure can distinguish whether the $A_\tau$-odd dynamics responsible for the polarization asymmetry occur in the fragmentation process, as in case(I) above, or in the target distribution, given by case(II) above. A similar selection can be performed in the case of associated baryon production in the Drell-Yan process. The possibility for this type of kinematic separation is important experimentally because the dynamics of spin-orbit effects in fragmentation are factorizable while in the effective distributions they are process dependent. Using the information described in Fig. 4 it is possible to explore spin-orbit dynamics in distributions in a systematic manner. We will return to a discussion of the mechanisms responsible for the generation of $k_{TN}$-asymmetries in Sec. IV with an illustration of the fractured Boer-Mulders effect for polarized $\Sigma\uparrow/\Lambda\uparrow$ production. At this point, it is appropriate to consider conjectured relationship between initial-state and final-state interactions in the Boer-Mulders functions and those in orbital distributions.

## III. The Parallelism Conjecture for Boer-Mulders Functions and Orbital Distributions

The parallelism conjecture for Boer-Mulders functions and orbital distributions was discussed in Ref. [15] in the context of the Georgi-Manohar [37] chiral quark model for spin-orbit dynamics in the proton. The most straightforward way to present the conjecture is this:

**Conjecture:** *The ratio of the Boer-Mulders function to the corresponding orbital distribution function is process independent.*

In terms of the nomenclature of Table 1, the conjecture can be written

$$\frac{\Delta^N G_{q\uparrow/p}(x, p_{TN}; Q^2)}{\Delta^N G_{q/p\uparrow}(x, p_{TN}; Q^2)} = r(x, p_{TN}; Q^2): \begin{array}{c} process \\ independent \end{array} \quad (3.1)$$

This conjecture embodies the idea that the initial-state and/or final-state interactions involved in the specification of the two $A_\tau$-odd distribution functions should cancel when the ratio is taken. If this connection can be established experimentally, it would indicate that both of these functions describe an intrinsic property of the proton even though each of them vanishes in the absence of initial-state or final-state interactions because of the underlying U(1) symmetry of the stable nucleon distributions.

For processes in the current fragmentation region the conjecture is of limited phenomenological value because, in the absence of final states with measurable polarization asymmetries it is difficult to test experimentally. However, the conjecture extends in a natural manner to the ratio of fractured functions,

$$\frac{\Delta^N M^q_{\Lambda\uparrow/\{q,q\}\uparrow:p}}{\Delta^N M^q_{\Lambda/(q,q):p\uparrow}} = r_\Lambda(x, p_{TN}; z, \vec{k}_T \cdot \vec{p}_T; Q^2): \begin{array}{c} process \\ independent \end{array} \quad (3.2)$$

that can be tested systematically since it possible to measure both target spin asymmetries and polarization asymmetries for the production of hyperons with parallel experimental situations. A measurement of process independence of the ratio in (3.2) would then provide support for the application of the same ratio for the more familiar functions given by (3.1).

The original conjecture given in (3.1) is suggested by an examination of Wilson operators involved in the measurement of the two distributions and also implies relationships between the twist-3 mechanisms leading to the related asymmetries in the limit of collinear factorization. It also predicts a change in sign for the Boer-Mulders function measured in the Drell-Yan process compared to that measured in SIDIS. The opportunities for checking the original conjecture are, however, limited. By contrast, the form of the conjecture given in (3.2) provides the basis for a systematic experimental program. Comparisons of target spin asymmetries with polarization asymmetries for hyperon production processes can be performed over a large range of kinematic variables with little experimental bias. In this context, we will now turn to a more thorough discussion of polarization asymmetries in the production of $\Lambda$'s and $\Sigma$'s.

### IV. The Production of Polarized $\Lambda$'s and $\Sigma$'s.

The fractured Boer-Mulders functions, $\Delta^N M^q_{B\uparrow/\{q,q\}\uparrow:p}(x, p_{TN}; z, \vec{p}_T \cdot \vec{k}_T; Q^2)$, can contribute to the polarization asymmetry of any baryon produced in the target fragmentation region of a hard process. As mentioned in the introduction, polarization asymmetries for hyperons have played a significant role in the history of spin physics because the weak decay of a hyperon allows direct access to information on its polarization. There exists a very large amount of data [1,2,28] on the polarization of hyperons in inclusive hadron-hadron and hadron-nucleus collisions. In this section, however, we are going to compare the production of polarized $\Lambda's$ with the production of polarized $\Sigma's$ in the target fragmentation region of semi-inclusive deep-inelastic lepton scattering, SIDIS, in order to illustrate the basic components of the fractured Boer-Mulders mechanism. We will see that the symmetry properties of axial vector diquarks lead to distinctly different predictions for the production of $\Lambda\uparrow's$ and $\Sigma\uparrow's$ by this mechanism. Simple kinematic arguments allow the sign of the fractured Boer-Mulders effect to be determined for these particles. The same arguments also indicate that the sign of the polarization changes for the fractured Boer-Mulders effect for hyperon production in SIDIS and associated production in the Drell-Yan process.

We can start this discussion by considering briefly the type of processes that can generate orbital angular momentum for diquarks within a stable nucleon. A natural initial approach to this question involves virtual transitions of the form

$$N\uparrow \to (N'\downarrow \pi)[L=+1]$$
$$N\uparrow \to (\Delta\uparrow \pi)[L=-1]$$
(4.1)

and

$$N\downarrow \to (N'\uparrow \pi)[L=-1]$$
$$N\downarrow \to (\Delta\downarrow \pi)[L=+1]$$
(4.2)

involving pionic fluctuations along with the transitions involving virtual kaons,

$$N\uparrow \to (\Sigma\downarrow K)[L=+1]$$
$$N\uparrow \to (\Sigma^*\uparrow K)[L=-1]$$
(4.3)

and

$$N\downarrow \to (\Sigma\uparrow K)[L=-1]$$
$$N\downarrow \to (\Sigma^*\downarrow K)[L=+1]$$
(4.4)

where we have assumed a broken SU(3) of flavor for baron structure. We only consider virtual processes involving nonzero orbital angular momentum, and, those virtual fluctuations in (4.1)-(4.4) that involve scalar diquarks are of no interest to us at this time since they do not contribute to fractured Boer-Mulders but only to fractured orbital distributions. However, let's examine a composite system of this type that contains an axial vector diquark, $\{q,q\}\uparrow$, and consider the correlation between orbital angular momentum and the "spin" orientation of the diquark. It is clear that all the fluctuations (4.1)-(4.4) lead to the correlation

$$\vec{J}_{\{q,q\}} \cdot \vec{L} = -1 \qquad (4.5)$$

independent of the initial spin of the stable nucleon. This strong correlation results from the overall conservation of angular momentum within the virtual transitions. The internal orbital angular momentum generated from such virtual processes is shared by all of the constituents in the strongly-interacting system. However, only particular dynamical configurations can generate a spin-directed momentum transfer associated with the orientation of diquark spin. The sketch shown in Fig. 5 illustrates an example of this type of configuration. This drawing indicates a possible virtual configuration of a nucleon involving a $\{q,q\}\uparrow$ diquark with $\vec{J}_{\{q,q\}} \cdot \hat{e}_y = +1$ and $\vec{L} \cdot \hat{e}_y = -1$. This quantum system displays a U(1) rotational invariance around the $\hat{y}$-axis, the orientation of the $\{q,q\}\uparrow$ spin. This rotational symmetry is necessarily broken by the hard scattering of a lepton from the quark <u>bound to the polarized diquark</u> within the virtual configuration. Because the scattered quark shares in the orbital angular momentum of the $\pi \otimes \{q,q\} \otimes q$ configuration, the hard scattering process at fixed Bjorken x preferentially occurs in the specific geometrical configuration in which the quark is moving toward the incoming lepton. The oriented force between the scattered quark and the polarized diquark then generates the spin-oriented momentum transfer, $\delta k_{TN}$, that determines the fractured Boer-Mulders effect as indicated in (2.11). A short "story board" of the scattering process that indicates the mechanisms involved in the generation of a transverse momentum asymmetry for a produced baryon is shown in two sketches of Fig. 6 indicating the role of the oriented confining force of the composite system in producing a polarization asymmetry.

This example shows that the <u>sign</u> of the fractured Boer-Mulders effect in SIDIS is determined by arguments similar to those applied by Burkardt [37] to determine the sign of the spin-dependent force leading to orbital distributions. Further discussion concerning oriented forces leading to single-spin asymmetries can be found in ref. [11]. Using the convention, discussed in Ref. [26], that chooses the positive $\hat{z}$-axis to be along the 3-vector $\vec{Q}$ defined by the hard scattering process, the polarization asymmetry is then <u>negative</u> for all baryons produced by the fractured Boer-Mulders effect.

The magnitude of the asymmetry for any specific baryon involves assumptions about the final-state interactions that require model-dependent calculations with ill-determined parameters. However, simple isospin arguments allow for a discussion of the relative magnitude of the fractured Boer-Mulders effect for $\Lambda\uparrow$ and $\Sigma\uparrow$ production. Consider the flavor content of the $\{q,q\}\uparrow$ diquarks generated in a nucleon target by the virtual pionic processes (4.1)-(4.2). The resulting diquarks must be from the isospin triplet

$$\{u,u\}\uparrow \quad \{u,d\}\uparrow \quad \{d,d\}\uparrow \tag{4.6}$$

Such diquark configurations cannot combine with an s quark generated during color fragmentation processes to produce a polarized $\Lambda\uparrow$. They can only combine with a strange quark to form an I=1 state, $\Sigma\uparrow$ or $\Sigma^*\uparrow$. The production of a polarized $\Lambda\uparrow$ in the fractured Boer-Mulders mechanism must involve the isospin doublet of diquarks

$$\{u,s\}\uparrow \quad \{s,d\}\uparrow \tag{4.7}$$

generated by the virtual fluctuations (4.3)-(4.4). When these diquarks join with u or d quarks during the fragmentation process they must form an equal mixture of I=0 $\Lambda\uparrow$ and I=1 $\Sigma\uparrow$ baryon states. The assumption that it is more probable to generate a strange quark in the fragmentation of a hard-scattering event than in the virtual transitions of a stable nucleon then leads to the ordering

$$\sum_{\{q,q\}} \Delta M^q_{\Sigma\uparrow/\{q,q\}\uparrow:p} \square \sum_{\{q,q\}} \Delta M^q_{\Lambda\uparrow/\{q,q\}\uparrow:p} \leq 0 \tag{4.8}$$

As discussed in Sec. II, the fractured Boer-Mulders functions and the polarizing fractured functions for baryons can be separately measured both in SIDIS and in associated production in the Drell-Yan process. Because this kinematic isolation is possible, it is of direct experimental interest to consider whether the fractured Boer-Mulders effect changes sign between the two processes in the manner of the Collins conjugation effect [29] predicted for orbital distributions. Simple arguments concerning the sign of the initial- state directed forces appropriate for baryon production in the Drell-Yan process sketched in Fig. 7 suggest strongly that the sign change occurs. The simple graphs shown in Fig. 8 then give a summary of the predictions for polarization asymmetries for $\Lambda\uparrow$ production and for $\Sigma\uparrow$ production both for the polarizing fractured functions of Ref. [26] and for the fractured Boer-Mulders effect as shown in (4.8) above. Experimental support for the pattern of polarization asymmetries indicated in Fig. 8 would help validate the connection between baryon polarization asymmetries [1,2,28] and the target spin asymmetries for meson production [16-20] discussed in the introduction. The change in sign between DY and SIDIS for these processes can be measured without the necessity of using polarized beams or polarized targets.

As indicated in (1.2) and (1.3) of the introduction, the experimental evidence for a mechanism leading to polarization in hyperon production that is <u>not</u> associated with the fragmentation process already exists from data on hadron-hadron collisions. If only fragmentation dynamics are considered, the flavor isospin of QCD combined with the factorization of $A_\tau$-odd effects in fragmentation functions cleanly predicts [26]

$$P(pN \to \Lambda_o \uparrow X) = -3P(pN \to \Sigma_o \uparrow X) \qquad (4.9)$$

and this ratio of polarizations is strongly contradicted by experimental results [28]. Previous efforts to find phenomenological explanations for hyperon polarization effects,[38.39], have not been able to explain this contradiction. The fractured Boer-Mulders effect considered here is the only theoretical mechanism for hyperon polarization in the target fragmentation region of hard processes that can explain the violation of the prediction (4.9) while being consistent with KPR factorization [26] and the results of Dharmarata and Goldstein [34]. The need for a theoretical connection between the mechanisms leading to polarization asymmetries in hyperon production and the mechanisms known to produce transverse target spin asymmetries in the current fragmentation region has recently been expressed by Aidala [40]. In [26] it was suggested that measurements of the polarizing fracture(d) functions can clarify and enhance the understanding of Collins functions. Here, we argue that measurements of the fractured Boer-Mulders functions can provide new insight into orbital distributions as well.

## V. Comparing Polarization Asymmetries and Target Spin Asymmetries for Baryon Production

Arguments concerning the origin of $A_\tau$-odd spin-directed momenta have previously been presented [15] that imply a close connection between those mechanisms within nonperturbative QCD leading to Collins functions and those leading to Boer-Mulders distributions and orbital distributions. Although the reasoning involved in these arguments is comparatively straightforward the existence of this important connection remains controversial and requires experimental support before it can be accepted. Experimental measurements of baryon spin asymmetries in the target fragmentation region of hard scattering processes can, in principle, provide a set of thorough and systematic tests of the interpretation of single-spin asymmetries in terms of the spin-orbit dynamics of QCD constituents that underlies the reasoning of Ref. [15]. Such experimental tests are not easy. Detector coverage sensitive to both the current fragmentation region and the target fragmentation region is required to separate $A_\tau$-odd mechanisms involving the fragmentation process from $A_\tau$-odd mechanisms associated with target distribution functions. Achieving this type of coverage in fixed target experiments can be a challenge and the study of baryon production asymmetries in such processes may best be suited to experiments at lepton hadron colliders.

Considering the target fragmentation region can augment our knowledge of nonperturbative dynamics because the interpretation of single-spin asymmetries in terms of the orbital angular momentum of SU(3) color constituents requires nontrivial internal correlations reflecting the confinement of the orbiting particles. In the fractured Boer-Mulders function, $\Delta^N M^q_{B\uparrow/\{q,q\}\uparrow:p}(x, p_{TN}; z, \vec{p}_T \cdot \vec{k}_T; Q^2)$, the confining chromodynamic force between an orbiting quark and diquark can, upon the hard scattering of the quark, produce an asymmetry in the transverse momentum of the attached axial-vector diquark that is associated with the orientation of the diquark's spin. These correlations are a consequence of the observation that orbital angular momentum within a confined composite system cannot be isolated onto an individual particle. In the drawings of Figs. 5 and 6, these correlations are indicated schematically by co-rotating "flux tubes". The sketches of flux tubes represent the localization of the inward-directed color electric force connecting the quark and diquark. This simple pictorial representation of the full correlation may be inadequate of incomplete. However, the consideration of a confining force as discussed in Sec. IV, does provide testable predictions for the relative magnitude of the polarizations of $\Lambda\uparrow$ and $\Sigma\uparrow$ associated with the fractured Boer-Mulders effect as well as a specific prediction for the sign of the polarization asymmetries in both SIDIS and in the associated production in the DY process. When combined with the predictions for the polarizing fracture(d) functions given in Ref. [26] we have found a comprehensive approach to baryon polarization asymmetries that both completes and enhances our understanding of other inclusive single-spin asymmetries.

Target spin asymmetries for the production of baryons in the target fragmentation region involve closely related dynamical mechanisms but do not rely on the parity-odd decay of hyperons to be experimentally observable. Therefore, there are many more processes that can be studied to provide important information about the flavor dependence of internal structure in hadrons. The fractured Collins-Heppelman functions, $\Delta^N M^q_{B/\{q,q\}\uparrow:p\uparrow}(x, p_T^2; z, k_{TN}; Q^2)$, can provide direct information concerning "transversity"- type effective diquark distribution functions. Such functions could not exist if diquarks, like gluons, were fundamental spin-1 constituents with negligible masses but represent allowed internal correlations in a multiparticle composite system. In contrast, the fractured orbital distributions, $\Delta^N M^q_{B/(q,q):p\uparrow}(x, p_{TN}; z, \vec{p}_T \cdot \vec{k}_T; Q^2)$, are sensitive to the orbital dynamics involving both scalar diquarks, $[q,q]$, as well as the spin-averaged contribution of axial-vector diquarks, $\{q,q\}\uparrow$. The separation of these two components of the nucleon structure in the production of different baryons in SIDIS follows arguments similar to those presented here. A further paper on target spin asymmetries for baryon production building on these ideas is currently in preparation.

**ACKNOWLEDGEMENTS**

The author has presented preliminary versions of this work to the Topical Workshop of the Hadron Physics interest group GHP 2009, Apr. 29-May 01, 2009, in Denver CO and to the Topical Workshop on Transverse Partonic Structure of Hadrons, June 21-26, 2009, in Yerevan, Armenia. Constructive comments from many participants from these workshops have been embodied in this manuscript. Special thanks are extended to Gerry Bunce and Harut Arvakyan.

*References*

1. G. Bunce, et al., Phys. Rev. Lett. **36**, 1113 (1976); K. Heller et al., Phys. Rev. **D16**, 2737 (1977); Phys. Lett. **B68**. 483 (1977)
2. L. Pondrom, Phys. Rep. **22**, 57 (1985)
3. G. Kane, J. Pumplin and W. Repko, Phys. Rev. Lett., **41**, 1683 (1978)
4. D. Sivers, Phys. Rev. **D41**, 83 (1990)
5. A.V. Efremov and O.V. Teryaev, Phys. Lett. **B150**, 303 (1985).
6. J. W. Qiu and G. Sterman, Phys. Rev. Lett. **67**, 2264 (1991) ;Phys. Rev. **D59**, 014004 (1998)
7. D. Sivers, Phys Rev. **D74**, 094008 (2006).
8. A. Artru, J. Czyzewski and H. Yabuki, Z. Phys. **C73**, 527 (1997)
9. R. D. Tangerman and P.J. Mulders, Phys. Rev. **D51**, 3357 (1995); Nucl. Phys. **B461**, 197 (1996).
10. J. Collins, Nucl. Phys. **B396**, 161 (1993).
11. D. Sivers, Phys. Rev. **D43**, 261 (1991).
12. D. Boer and P. Mulders, Phys. Rev. **D57**, 5780 (1998)
13. A. Baccheta, U. D'Alesio, M. Diehl and C.A. Miller, Phys. Rev. **D70**, 117504 (2004)
14. D. Sivers, *Proceedings of Advanced Research Workshop on High Energy Spin Physics (DSPIN-07)* A.V. Efremov and S.V. Goloskokov editors, (JINR press, Dubna, 2008) p. 161
15. D. Sivers, arXiv:0704.1791 (2007) submitted to Phys. Rev. D but rejected.
16. See, for example, A. Airapetian et al. (for the Hermes collaborration), Phys. Rev. Lett. **94**, 012003
17. See, for example, E.S. Ageev et al., (for the Compass experiment). Nucl. Phys. **B765**, 31 (2007)
18. A. Afanasev, et al., arXiv:hep-ph/0703288 (JLAB)
19. J. Adams et al. (Star Collaboration) Phys. Rev. Lett. **92**, 171801 (2004); B.I. Abelev et al. (Star Collaboration) Phys. Rev. Lett. **101**, 222001 (2008)
20. H. Liu, (for the Phenix Collaboration), in *Spin Physics, 18$^{th}$ Int. Spin Physics Symposium*, (Editors D.G. Crabb et al., AIP, New York, 2009) p. 439
21. J. Collins, S. Heppelmann and G. Ladinski, Nuc. Phys. **B420**, 565 (1994).
22. R. Seidl, et al. ( Belle collaboration), Phys. Rev. Lett. **96**, 232002 (2006

23. M. Anselmino, et al, Phys. Rev. **D75**, 054032 (2007)
24. L. Trentadue and G. Veneziano, Phys. Lett. **B323**, 201 (1994)
25. F. Ceccopieri and L. Trentadue, Phys. Lett. **B660**, 43 (2008); **B655**,15 (2007)
26. D. Sivers, Phys. Rev. **D79**, 085008 (2009)
27. B. Andersson er al., Phys. Lett. **B85**, 417, (1979); Phys. Rep. **97**, 31 (1983).
28. K. Heller, Proceedings of Spin 96; C.W. de Jager, T.J. Kobel and P. Mulders, Eds. (World Scientific, NY, 1997); A.D. Panagiotou, Int. J. Mod. Phys. **A5**,1197 (1990)
29. J. Collins, Phys. Lett B **536**, 43 (2002).
30. R.L. Jaffe, Phys. Rep. **409**, 1 (2005); Phys. Rev. **D72**, 074508 (2005)
31. S.J. Brodsky, D.S. Huang and I. Schmidt, Phys. Lett. **B530**,99 (2002) ; L.P. Gamberg, et al., Phys. Rev. **D67**, 071504 (2003).
32. For an example of factorization prescription issues in heavy quark DIS, see S. Alekhin et al. arXiv:0908.312 (2009)
33. J. Collins and J.-W. Qiu, Phys. Rev. **D75**, 114014 (2007); W. Vogelsang and F. Yuan, Phys. Rev. **D76**, 094013 (2007).
34. W. G. Dharmaratna and G. Goldstein, Phys. Rev. **D41**, 1731 (1990); Phys. Rev. **D53**, 1073 (1996)
35. Y. Koike and K. Tanaka, Phys. Lett. **B 646**, 232 (2007); arXiv0907.2797
36. A Bacchetta, C. Bomhof, P. Mulders and F. Pijlman, Phys. Rev. **D72**, 034030 (2005)
37. M. Burkardt, Phys. Rev. **D66**, 114005 (2002); Phys. Rev. **D69**, 057501 (2004)
38. T. DeGrand and H. Miettenen, Phys. Rev. **D23**, 227 (1981); Phys. Rev. **D24**, 2419 **(1981)**
39. J. Szwed, Phys. Lett. **B105**, 403 (1981)
40. C. Aidala, in *18$^{th}$ International Spin Physics Symposium*, editors D.G. Crabb, et al. (American Institute of Physics, NY 2009) p. 124

$$\Delta^N D_{h/q\uparrow}(z,k_{TN}) = \frac{2k_{TN}}{zM_h} H_1^{Tq}(z,k_{TN}) \qquad \Delta^N D_{h\uparrow/q}(z,k_{TN}) = \frac{k_{TN}}{zM_h} D_1^{Tq}(z,k_{TN})$$

*Collins_functions*          *polarizing_fragmentation_functions*

$$\Delta^N G_{q\uparrow/p}(x,p_{TN}) = -\frac{p_{TN}}{M_p} h_1^{Tq}(x,p_{TN}) \quad \Delta^N G_{q/p\uparrow}(x,p_{TN}) = -\frac{2p_{TN}}{M_p} f_1^{Tq}(x,p_{TN})$$

*Boer–Mulders_functions*          *orbital_structure_functions*

Table 1.
This table gives the relationship between the expressions used by the author for partonic number densities in terms of the expressions commonly used for the related correlators that have the dimension of inverse momentum.

$$\Delta^N M_{h/p\uparrow}^{q\uparrow}(x,p_T^2;z,k_{TN};Q^2) \qquad\qquad \Delta^N M_{h\uparrow/p}^{q}(x,p_T^2;z,k_{TN};Q^2)$$

*Collins–Heppelmann_fracture_functions*    *polarizing_fracture_functions*

$$\Delta^N M_{h\uparrow/p}^{q\uparrow}(x,p_{TN};z,\vec{k}_T\cdot\vec{p}_T;Q^2) \quad \Delta^N M_{h/p\uparrow}^{q}(x,p_{TN};z,\vec{k}_T\cdot\vec{p}_T;Q^2)$$

*Boer–Mulders_fracture_functions*    *orbital_fracture_functions*

Table 2.
Transverse-momentum dependent fracture functions that characterize the $A_\tau$-odd single-spin asymmetries in the current fragmentation region of semi-inclusive deep inelastic lepton scattering are classified in this table. The fracture function formalism defines conjoint probability distributions without the requirement of assuming hard-scattering factorization.

$$\Delta^N M^q_{B/\{q,q\}\uparrow:p\uparrow}(x,p_T^2;z,k_{TN};Q^2) \qquad \Delta^N M^q_{B\uparrow/(q,q):p}(x,p_T^2;z,k_{TN};Q^2)$$

*fractured _ Collins – Heppelmann _ functions    polarizing _ fracture(d) functions*

$$\Delta^N M^q_{B\uparrow/\{q,q\}\uparrow:p}(x,p_{TN};z,\vec{k}_T\cdot\vec{p}_T;Q^2) \quad \Delta^N M^q_{B/(q,q):p\uparrow}(x,p_{TN};z,\vec{k}_T\cdot\vec{p}_T;Q^2)$$

*fracturedBoer – Mulders _ functions    fractured _ orbital _ functions*

Table 3.

The fractured functions in this table describe mechanisms for $A_\tau$-odd single-spin asymmetries in the production of baryons in the target fragmentation region of semi-inclusive deep inelastic scatterings. These fractured functions specify effective distribution functions for diquarks that mirror the effective distribution functions for quarks classified in Table 2.

**FIGURE CAPTIONS**

Fig. 1. A semi-inclusive lepton scattering process from an unpolarized ensemble of protons is sketched in this figure. The final state includes a quark jet and a baryon in the target fragmentation region associated with a $\{q,q\}\uparrow$ diquark. The 3-momentum of the produced baryon lies in the x-z plane and the spin state is characterized by a density matrix that is diagonal when the spin is quantized in the y direction. The $A_\tau$-odd spin-directed momentum leading to a polarization asymmetry for the produced baryon can originate in coherent spin-orbit dynamics in the target distribution functions (to the left of the first horizontal line) or in the coherent spin-orbit dynamics (to the right of the second horizontal line) associated with the fragmentation of the $\{u,d\}\uparrow$ diquark into the final-state baryon. No $A_\tau$-odd dynamical effects are generated directly by the hard-scattering process (the cross-hatched region between the two horizontal lines) since light-quark QCD perturbation theory obeys Kane, Pumplin, Repko [3], (KPR), factorization. The local hard-scattering process, however, doest break the rotational U(1) invariance of the virtual spin-orbit fluctuations in the stable nucleon. This symmetry breaking leads to process dependence in the $A_\tau$-odd "effective distributions" generated in the initial state.

Fig. 2   The convolutions in transverse momentum leading to the transverse momentum of a final-state baryon in the target fragmentation region is shown in the foreground while those convolutions leading to the transverse momentum for a quark jet in the current fragmentation region are shown in the background.  Following the divisions of Fig. 1, we start with the intrinsic transverse momenta of the diquark (foreground) and quark (background).  These are modified by radiative transverse momentum generated during the hard-scattering process.  In addition, a component of the transverse momentum of the final-state baryon is also generated in the fragmentation process. The momenta of hadrons in the current fragmentation region are summed to specify the kinematics of the quark jet.  Transverse momenta of central particles are also indicated.

Fig. 3   Based on the arguments in the text, the possibilities for producing an $A_\tau$-odd spin-directed momentum leading to a polarization asymmetry in baryon production are shown in separate diagrams.  The fractured Boer-Mulders effect has $A_\tau$-odd dynamics associated with the distribution function while the polarizing fractured effect generates $A_\tau$-odd momentum transfer in the fragmentation process.

Fig. 4   The kinematic separation of the fractured Boer-Mulders functions and the polarizing fractured functions in SIDIS involved weighting experimental events by the angles defining the orientation of the planes shown in this drawing.  All three planes share the same positive $\hat{z}$-axis that is defined to align with the spacelike momentum transfer $\vec{Q}$ of the hard-scattering process.  The angles between the positive $\hat{x}$-axes for the planes are defined to be $\phi_{lB}, \phi_{lq}, \phi_{qB}$ with $\phi_{lB} = \phi_{lq} + \phi_{qB}$.  The polarization asymmetry for the produced baryon must be an odd function of $\phi_{lB}$.  The fractured Boer-Mulders effect producing the asymmetry is odd in $\phi_{lq}$ and even in $\phi_{qB}$ while the polarizing fractured effect is even in $\phi_{lq}$ and odd in $\phi_{qB}$.

Fig. 5   A sketch indicating geometric correlations in a virtual fluctuation of a nucleon into an $L_y = -1$ quark diquark pion configuration displays a directed force between the orbiting quark and diquark indicated schematically by a co-rotating chromoelectric flux tube.

Fig. 6   Drawings suggesting the origin of spin-directed momentum in the fractured Boer-Mulders effect for SIDIS are shown in these two sketches.  At fixed Bjorken-x, lepton quark scattering occurs preferentially from a quark rotating <u>toward</u> the incoming lepton.  This situation is indicated for an $L_y = -1$ virtual $\pi \otimes \{q,q\}\uparrow \otimes q$ configuration in the top panel.  The bottom panel shows the confining force (indicated by a co-rotating flux tube) producing a negative $\partial p'_{TN}$ for the localized axial-vector diquark and a positive $\partial p_{TN}$ for the scattered quark.

Fig. 7   In the fractured Boer-Mulders effect for the associated production of baryons in the Drell-Yan process, the quark antiquark annihilation preferentially occurs when the quark in the target nucleon is rotating <u>toward</u> the incoming beam.  The annihilation

releases energy from the confining flux resulting in the recoil of the polarized axial-vector diquark with positive $\partial p'_{TN}$ while the produced lepton pair exhibits negative $\partial p_{TN}$.

Fig.8  Graphs comparing the predictions for polarization asymmetries of $\Lambda_o \uparrow$ and $\Sigma_o \uparrow$ produced in SIDIS and in Associated Drell-Yan are shown for discussion. The polarizing fracture functions (top panels) factorize in the two processes. The ratio $P(\Lambda_o \uparrow)/P(\Sigma_o \uparrow) = -3$ results from isospin C.G. coefficients in the fragmentation process. The fractured Boer-Mulders functions are negative in SIDIS and positive in Drell-Yan for both hyperons. In each case the magnitude is greater for $\Sigma_o \uparrow$ than for $\Lambda_o \uparrow$ because of the contribution of I=1 axial-vector diquarks.

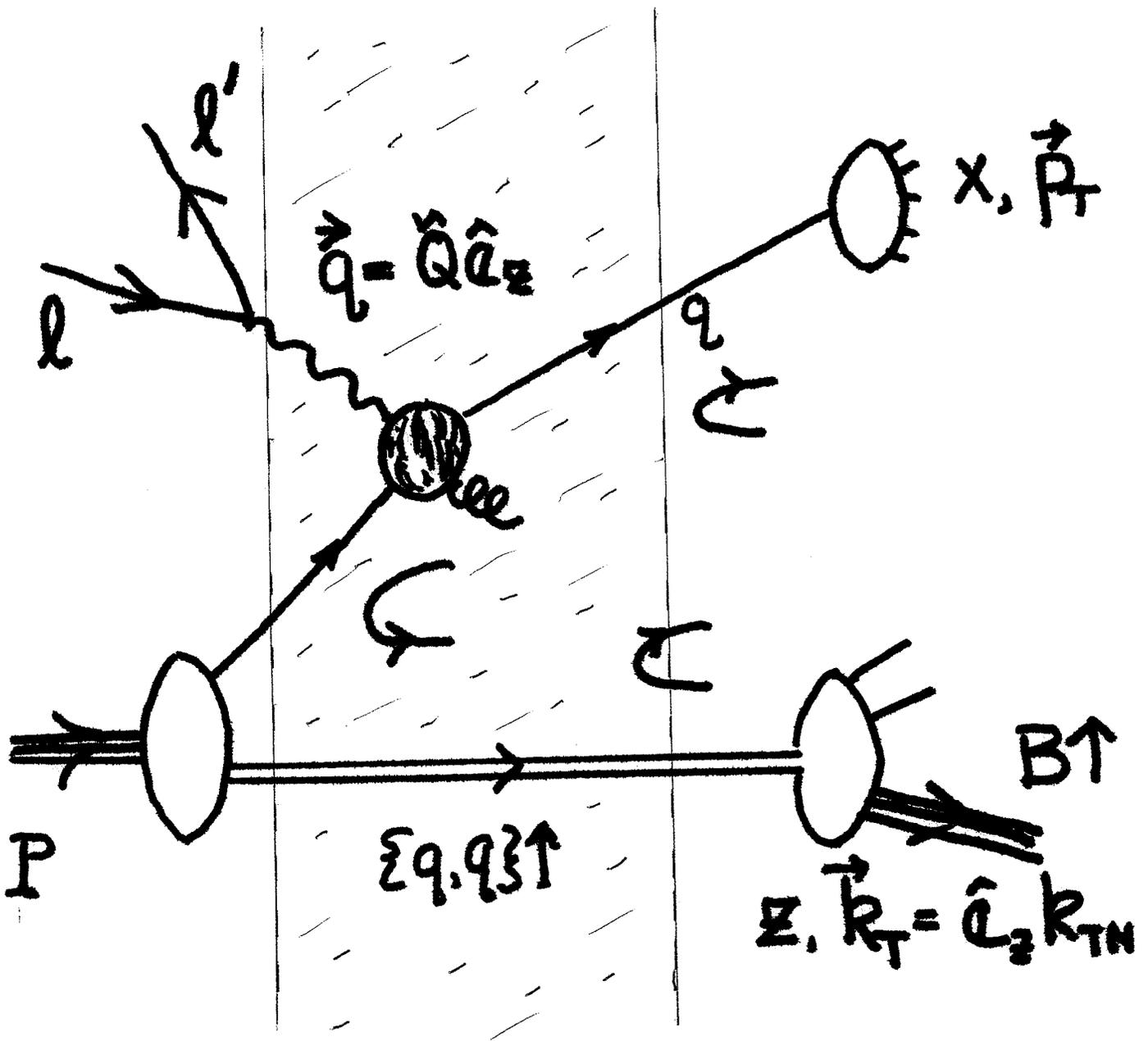

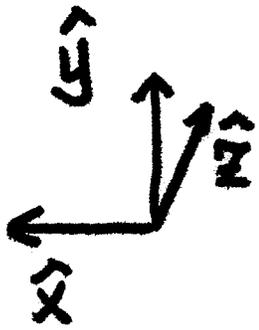

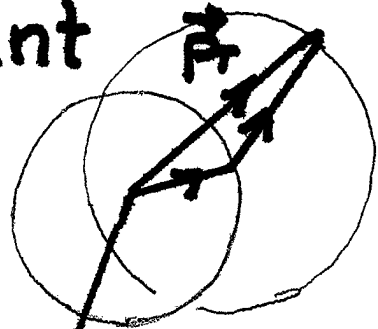

current

$\vec{p}_T$

transverse
momenta

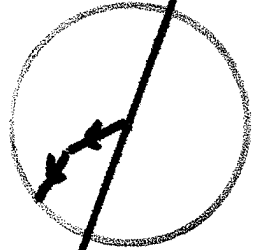

target

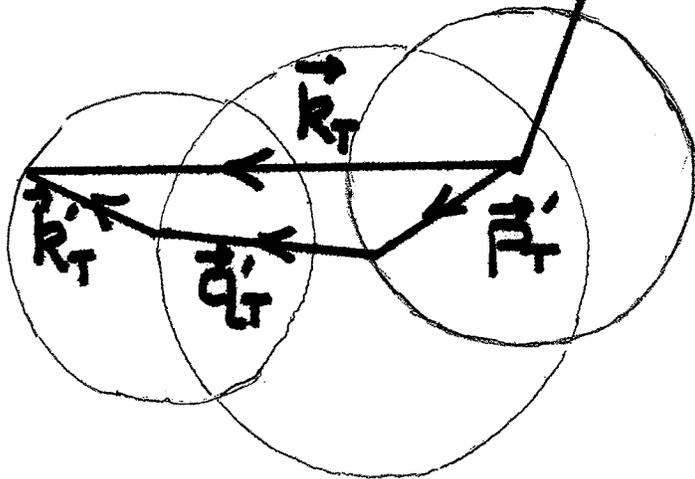

$$\vec{k}_T = k_{TM}\hat{a}_x = \vec{p}_T' + \vec{q}_T' + \vec{k}_T'$$

# Spin-Oriented Momentum

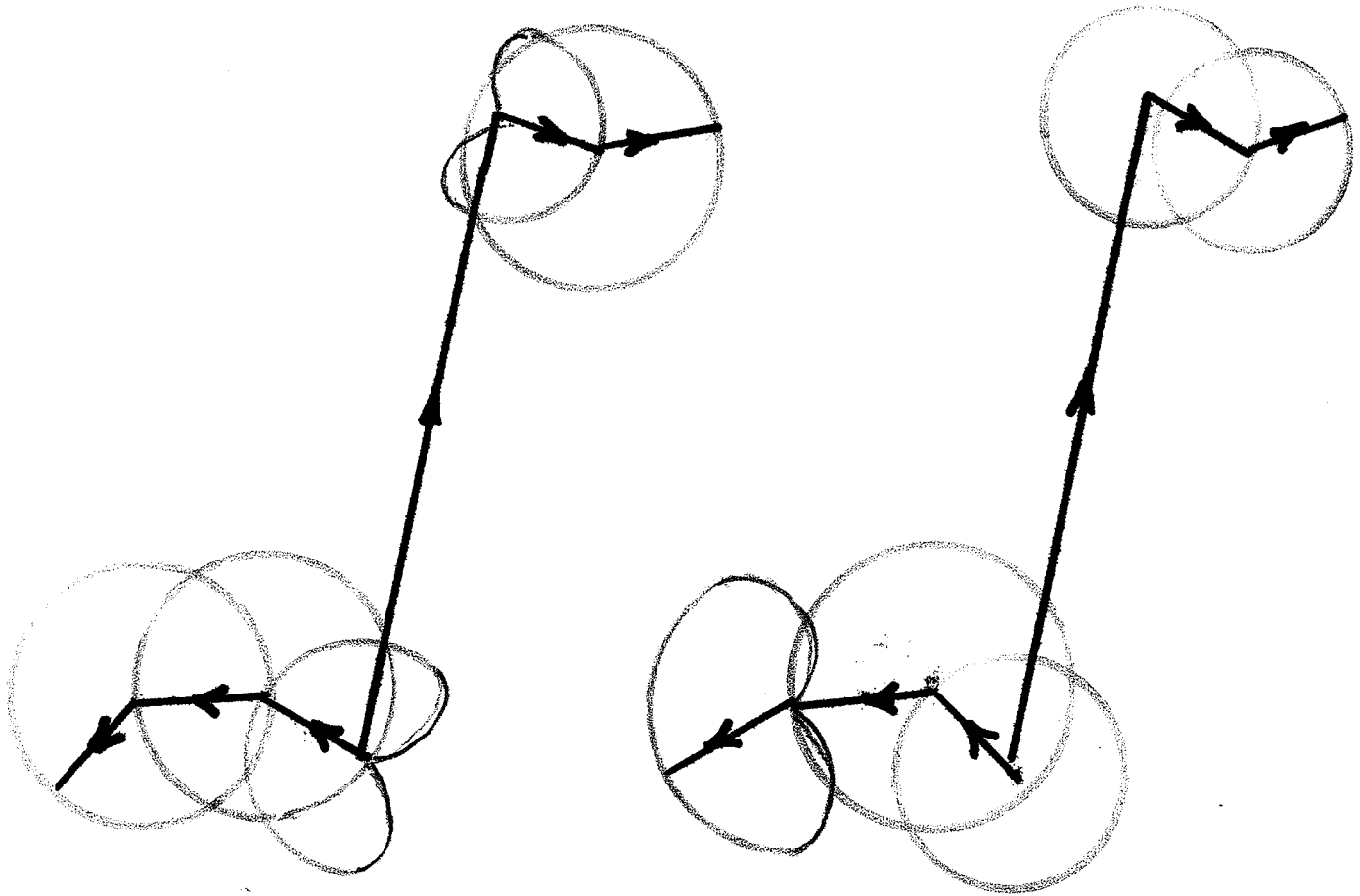

Fractured
Boer-Mulders
$A_T$·odd in
distribution

]   [

Polarizing
Fractured function
$A_T$·odd in
fragmentation

# SIDIS

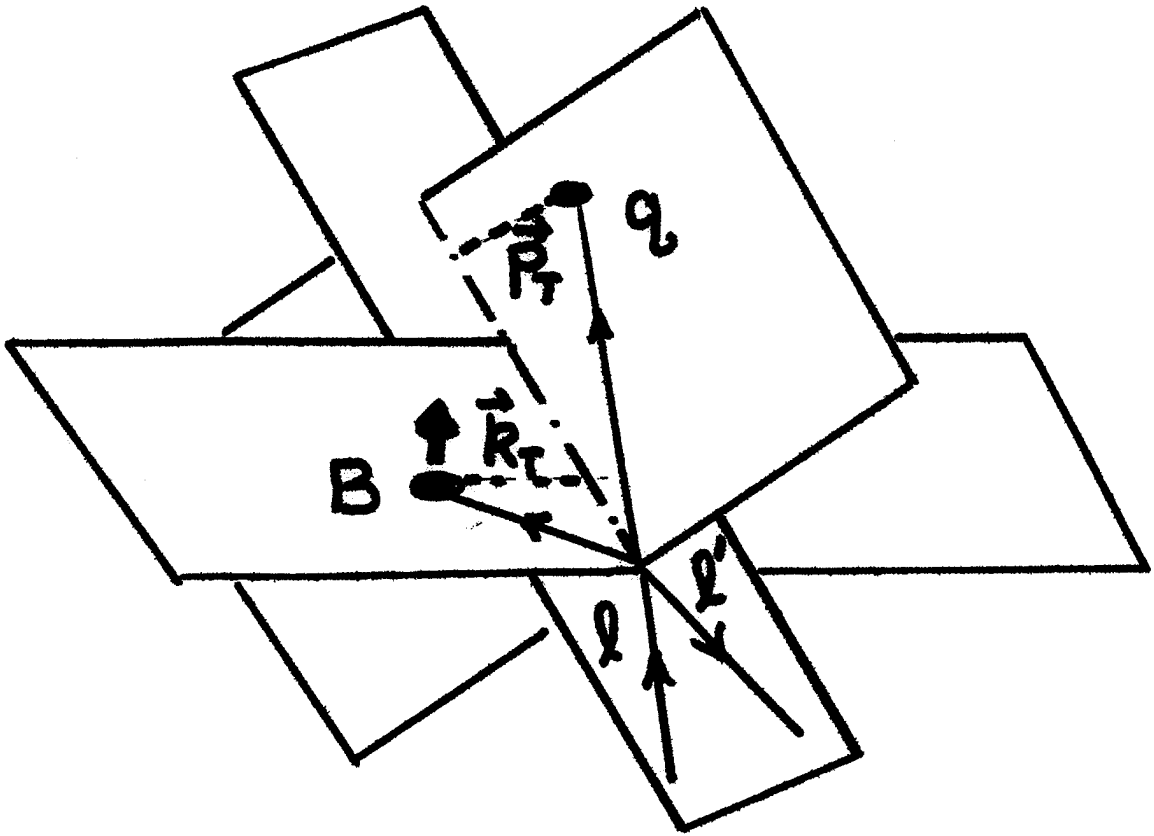

## Three Planes

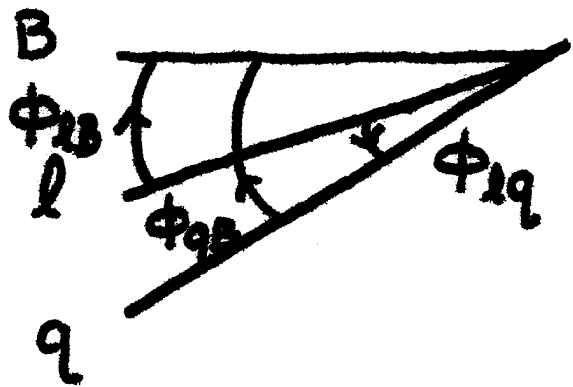

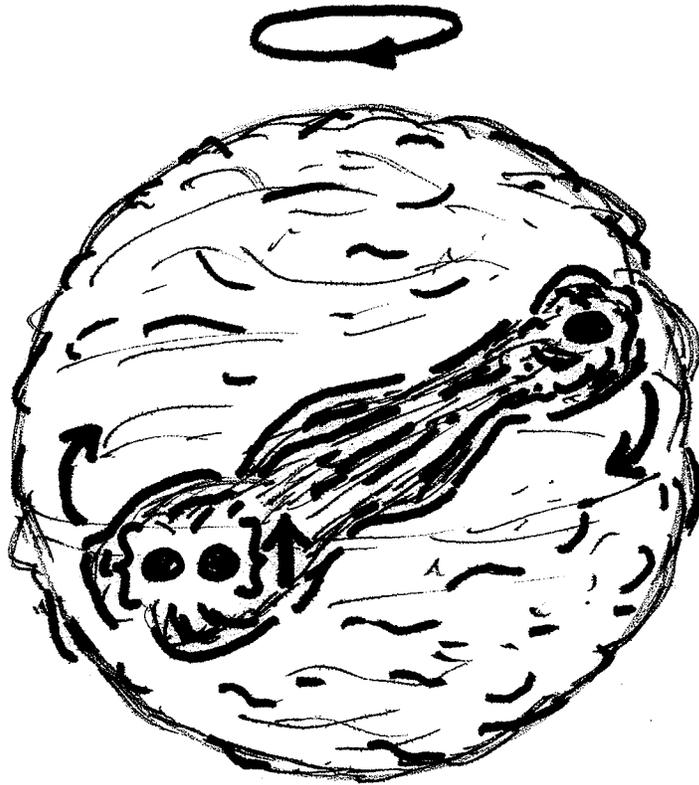

$$(\pi \bullet \{q,q\} \uparrow \bullet q) \mathbf{L}_y = -1$$

Spin · Orbit Dynamics

# SIDIS

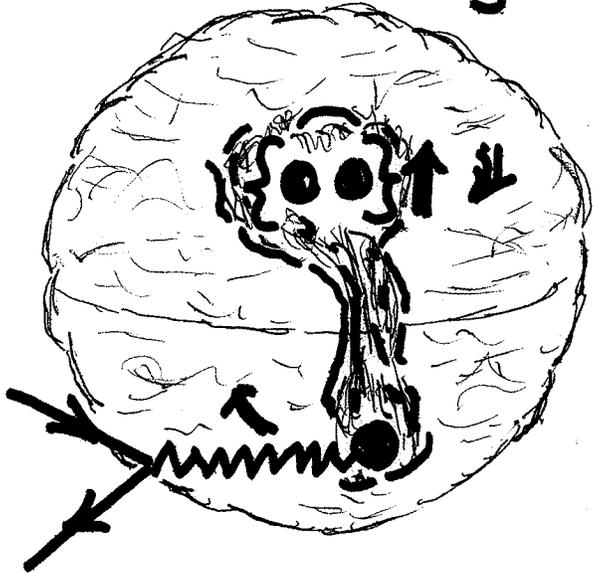

$L_y = -1$

## hard scattering

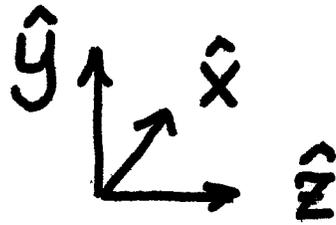

$\hat{y}$ $\hat{x}$ $\hat{z}$

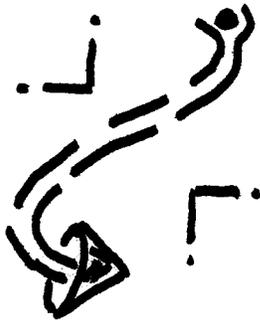

spin-directed force

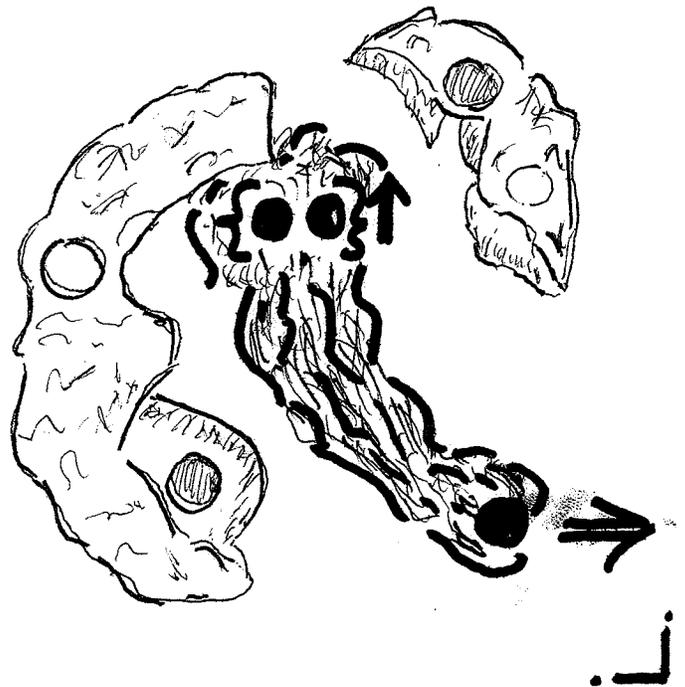

# Associated Drell-Yan

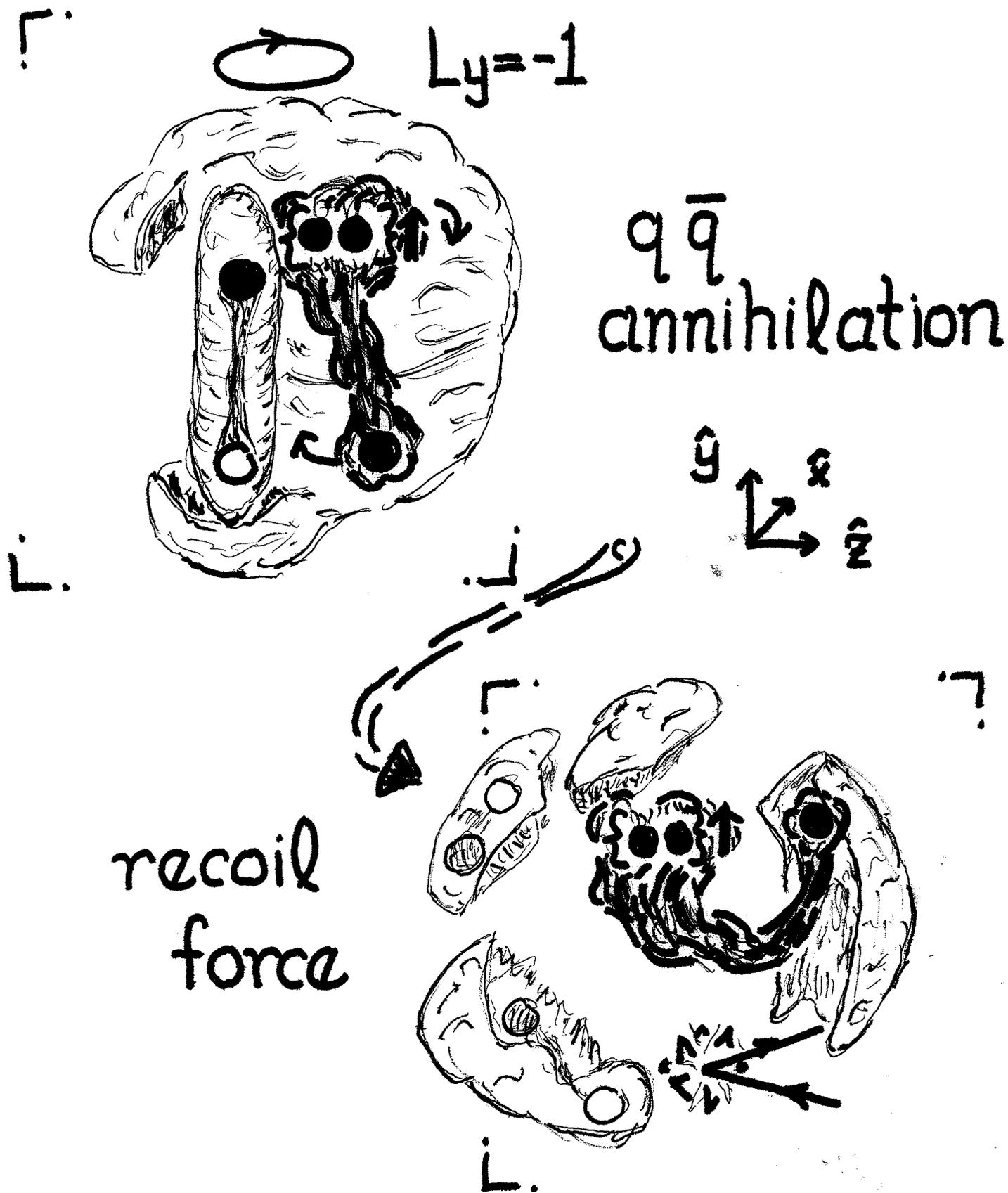

$L_y = -1$

$q\bar{q}$ annihilation

recoil force

## Polarizing Fract.

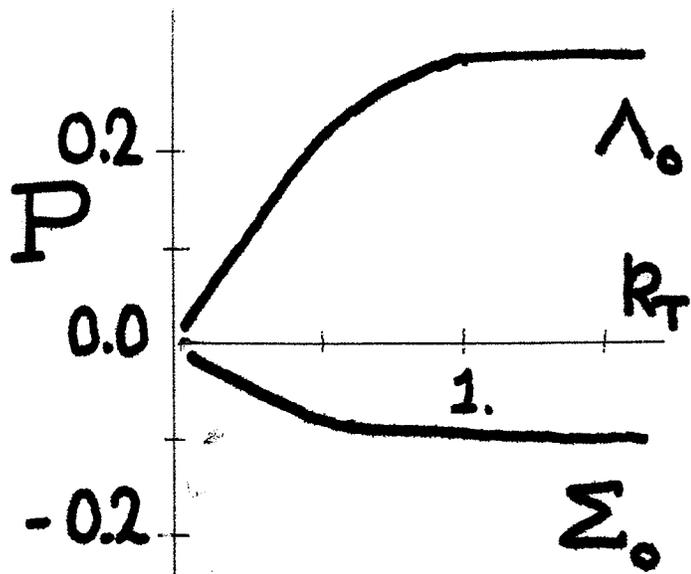

SIDIS

## Polarizing Fract.

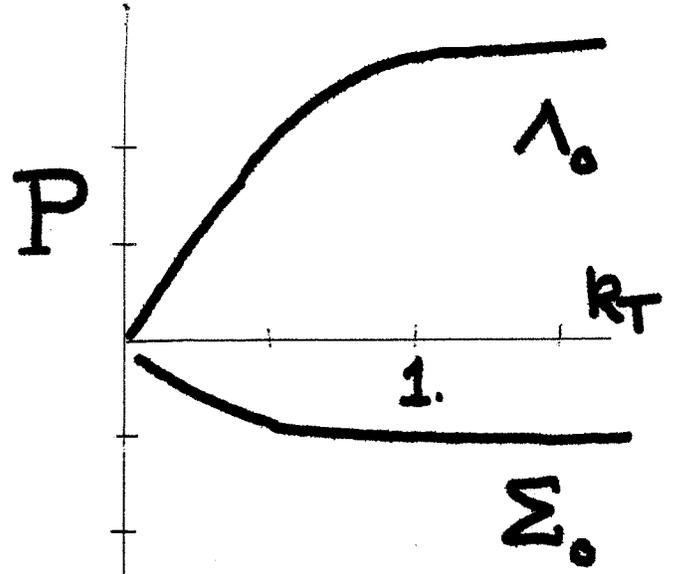

Ass. D.-Y.

## Fract. Boer-Mulders

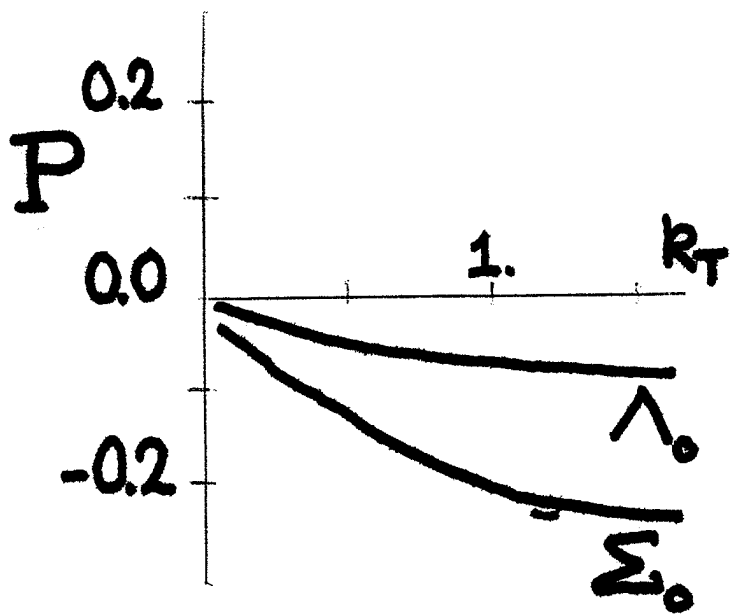

SIDIS

## Fract. Boer-Mulders

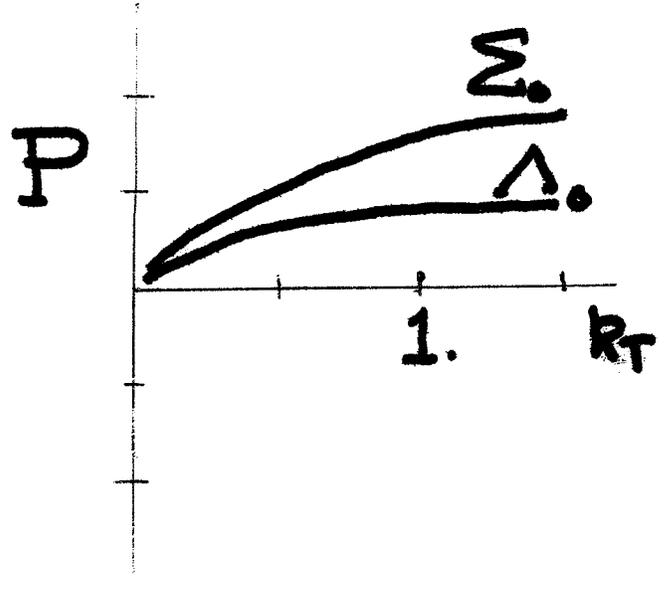

Ass. D.-Y.

8